\documentclass[11pt]{article}

\usepackage{graphicx}
\usepackage{citesort}
\usepackage{amssymb}
\usepackage{amsfonts}
\usepackage{amsmath}
\usepackage{epsfig}
\usepackage{color}
\usepackage{fancybox}
\usepackage{textcomp}
\usepackage{multirow}
\usepackage{appendix}
\usepackage{setspace}
\usepackage{psfrag}
\usepackage[ruled,vlined,linesnumbered]{algorithm2e}

\renewcommand{\baselinestretch}{1.35}
\newtheorem{Lemma}{Lemma}

    \def\ang#1{\mbox{$\langle #1 \rangle$}}

    \def\Complex{{\rm\rule[.23ex]{.03em}{1.1ex}\kern-.3em{C}}}

    \newcommand{\be}{\begin{equation}} \newcommand{\ee}{\end{equation}}
    \newcommand{\bea}{\begin{eqnarray}} \newcommand{\eea}{\end{eqnarray}}
    \newcommand{\benum}{\begin{enumerate}} \newcommand{\eenum}{\end{enumerate}}

    \newcommand{\qb}{{\bf b}}

    \newcommand{\qe}{{\bf e}}

    \newcommand{\qs}{{\bf s}}

    \newcommand{\qx}{{\bf x}}
    \newcommand{\qy}{{\bf y}}
    \newcommand{\qz}{{\bf z}}

    \newcommand{\qA}{{\bf A}}
    \newcommand{\qB}{{\bf B}}
    \newcommand{\qC}{{\bf C}}
    \newcommand{\qD}{{\bf D}}
    \newcommand{\qE}{{\bf E}}
    \newcommand{\qF}{{\bf F}}
    
    \newcommand{\qH}{{\bf H}}
    \newcommand{\qI}{{\bf I}}

    \newcommand{\qR}{{\bf R}}
    
    \newcommand{\qT}{{\bf T}}
    
    \newcommand{\qV}{{\bf V}}
    \newcommand{\qW}{{\bf W}}
    \newcommand{\qX}{{\bf X}}

    \newcommand{\qzero}{{\bf 0}}

    \newcommand{\qwbE}{\overline{\bf E}}
    \newcommand{\qwbF}{\overline{\bf F}}
    \newcommand{\qwbT}{\overline{\bf T}}
    \newcommand{\qwbR}{\overline{\bf R}}

    \newcommand{\qDelta}{{\boldsymbol \Delta}}

    \newcommand{\qSigma}{{\boldsymbol \Sigma}}

    \newcommand{\qOmega}{{\boldsymbol \Omega}}

    \newcommand{\qnu}{{\boldsymbol \nu}}
    \newcommand{\qmu}{{\boldsymbol \mu}}

    \newcommand{\bbB}{{\mathbb B}}

    \newcommand{\bbC}{{\mathbb C}}
    
    \newcommand{\bbU}{{\mathbb U}}

    \newcommand{\tr}{{\sf tr}}
    \newcommand{\Ex}{{\sf E}}

    \newcommand{\sfZ}{{\sf Z}}

\vspace{-0.5in}
\title{Message Passing Algorithm for Distributed Downlink Regularized Zero-forcing Beamforming with Cooperative Base Stations}

\author{Chao-Kai Wen\footnote{Institute of Communications Engineering, National Sun Yat-sen University, Taiwan. {\sf Email: chaokai.wen@mail.nsysu.edu.tw}.},
Jung-Chieh Chen\footnote{The Department of Optoelectronics \& Communication Engineering, National Kaohsiung Normal University, Kaohsiung, Taiwan.},
Kai-Kit Wong\footnote{Department of Electronic and Electrical Engineering, University College London, UK.},
and~Pangan Ting\footnote{Industrial Technology Research Institute (ITRI), Hsinchu 310, Taiwan, R.O.C.}}

\date{}

\begin{document}
\maketitle

\begin{abstract}
Base station (BS) cooperation can turn unwanted interference to useful signal energy for enhancing system performance. In the cooperative downlink, zero-forcing
beamforming (ZFBF) with a simple scheduler is well known to obtain nearly the performance of the capacity-achieving dirty-paper coding. However, the centralized
ZFBF approach is prohibitively complex as the network size grows. In this paper, we devise message passing algorithms for realizing the regularized ZFBF (RZFBF) in
a distributed manner using belief propagation. In the proposed methods, the overall computational cost is decomposed into many smaller computation tasks carried
out by groups of neighboring BSs and communications is only required between neighboring BSs. More importantly, some exchanged messages can be computed based on
channel statistics rather than instantaneous channel state information, leading to significant reduction in computational complexity. Simulation results
demonstrate that the proposed algorithms converge quickly to the exact RZFBF and much faster compared to conventional methods.
\end{abstract}

{\bf Index Terms}---Base station cooperation, Belief-propagation, Distributed algorithm, Message passing, Zero-forcing beamforming.

\newpage
\section*{\sc I. Introduction}
Multiuser multiple-input multiple-output (MU-MIMO) antenna system has been recognized as an effective means to increase capacity in the downlink
\cite{Spencer-04COMMag,ZGPan-04,Gesbert-07SigMag}. However, MU-MIMO may not be as effective if edge-of-cell users are concerned due to the severe inter-cell interference that is hard to
suppress. In recent years, it has emerged that letting base stations (BS) cooperate can greatly improve the link quality of the edge-of-cell users by turning unwanted interference into useful
signal energy, e.g., \cite{Shamai-97IT,Caire-03IT,Somekh07IT,Somekh09IT,Gesbert10JSAC,Akai-Kit-10} (and the references therein). Ideally, by sharing all the required information via high-speed
backhaul links, all BSs in a downlink cellular network can become a super BS with distributed sets of antennas. This architecture will then allow the use of well-known optimal or suboptimal
transmission strategies such as capacity-achieving dirty-paper coding (DPC) techniques \cite{Costa-83IT,Caire-03IT,Somekh07IT} and zero-forcing beamforming (ZFBF) \cite{Yoo-06JSAC,Somekh09IT},
respectively.

Although DPC is capacity-achieving, it is very complex and massive interest has been to employ ZFBF with a simple scheduler to approach near-capacity performance \cite{Yoo-06JSAC,Somekh09IT}.
For example, several testbeds for implementing BS cooperation have adopted ZFBF techniques, e.g., \cite{Sam-07,Irm-09,Jun-10,Irm-11}. Regularized ZFBF (RZFBF) is a generalization of ZFBF by
introducing the regularization parameter \cite{Joham-02ISSSTA,Peel-05Tcom}. It has been revealed that several beamformers can have a RZFBF structure by selecting the regularization parameter
properly \cite{Zak-12IT}. Even though information-theoretic studies have provided overwhelming support to RZFBF \cite{Simeone-11BOOK,Zak-12IT,Hoydis-12JSAC}, the real question is how could RZFBF
be implemented in a very large-scale cellular network?

A straightforward way to implement RZFBF would be to require that there is a central processing unit which possesses all the necessary channel state information
(CSI) and performs the entire optimization. However, as a network expands with more BSs cooperating, it becomes inviable to perform joint processing over all BSs
because of the limiting backhaul capacity and the excessive computational complexity. It is therefore of greater interest to consider an architecture where BSs
only communicate with neighboring BSs and the overall computation cost is decomposed into many smaller computational tasks, amortized by groups of smaller number
of cooperating BSs. Motivated by this, in this paper, we propose two message passing algorithms to realize RZFBF in a distributed manner. The proposed approaches
are particularly well suited to cooperation of large clusters of simple and loosely connected BSs. Most importantly, in our designs, each BS is only required to
know the data symbols of users within its reception range rather the entire cellular network, greatly reducing the backhaul requirements.

The use of distributed methods in beamforming computations has been studied recently in
\cite{Ng-08IT,Aktas-11BOOK,Bjornson-10SP,Sohn-11TWC,Huang-11SP,Rangan-12JSAC}. Our approach is similar to \cite{Ng-08IT} in that both aim at achieving RZFBF and
use belief propagation (BP). Nonetheless, the two approaches differ considerably. Our main contributions are summarized as follows:

\begin{itemize}
\item First, we generalize the earlier results in \cite{Ng-08IT} to incorporate \emph{multiple antennas}
at both BSs and user equipments (UEs) and our results can be applied to a wide range of scenarios with \emph{complex-valued systems}. Further, we adopt the
approximate message passing (AMP) method in \cite{Donoho-09PNAS} to significantly reduce the number of exchange messages. The proposed AMP-RZFBF exhibits the
advantage that every communication of BS with its neighbors only takes place in a broadcast fashion as opposed to the unicast manner in \cite{Ng-08IT}. The
used AMP method has recently received considerable interest in the field of compressed sensing \cite{Donoho-09PNAS,Bayati-11IT,Krzakala-12JSM,Rangan-10ArXiv}.
Our form of the message passing algorithm is closely related to the AMP methods in \cite{Krzakala-12JSM} which are a special case of the generalized AMP
\cite{Rangan-10ArXiv}.

\item In AMP-RZFBF, BSs must compute several matrix inversions for every channel realization and then exchange these auxiliary parameters among themselves, requiring very high computational capability and rapid information exchange between the BSs. To tackle this, we approximate some of the auxiliary parameters by exploiting the spatial channel covariance information (CCoI). The CCoI-aided AMP-RZFBF results in significantly
simpler implementations in terms of computation and communication. With the CCoI-aided AMP-RZFBF, the BSs compute and exchange the auxiliary parameters at the time scale merely at which the
CCoI changes but not the instantaneous CSI. Simulation results show that CCoI-aided AMP-RZFBF achieves promising results, which are different from earlier results based on the CCoI, e.g.,
\cite{Hoydis-11ISIT,Lakshminarayana-12SP}, where a performance degeneration is usually expected.

\item Implementing RZFBF in a distributed manner can be achieved by an optimization technique called the alternating direction method of multipliers (ADMM) approach in \cite{Boyd-11BOOK}. Applications of ADMM to the concerned beamforming problem can be found in \cite{Zhu-10TWC} (or \cite[Section 8.3]{Boyd-11BOOK}). However, it is known that ADMM can be very slow to converge. Simulation results will demonstrate that our proposed message passing algorithms exhibit a much faster convergence rate when compared to ADMM.
\end{itemize}

{\em Notations}---Throughout this paper, the complex number field is denoted by $\mathbb{C}$. For any matrix $\qA\in \bbC^{M \times N}$, $A_{ij}$ denotes the $(i,j)$th entry, while $\qA^T$, and
$\qA^H$ return the transpose and the conjugate transpose of $\qA$, respectively. For a square matrix $\qB$, $\qB^\frac{1}{2}$, $\qB^{-1}$, ${\sf tr}(\qB)$, and $\det(\qB)$ denote the principal
square root, inverse, trace, and determinant of $\qB$, respectively. In addition, $\qI_N$ is an $N \times N$ identity matrix, $\qzero_N$ denotes either an $N \times N$ zero matrix or a zero
vector depending on the context, and $\qe_i$ denotes the column vector with the $i$th element being $1$ and $0$ elsewhere. Finally, $\|\cdot\|_2$ represents the Euclidean norm of an input
vector, and $\Ex \{\cdot\}$ returns the expectation of an input random entity.

\section*{\sc II. System Model and Problem Formulation}
\begin{figure}
\begin{center}
\resizebox{5.5in}{!}{%
\includegraphics*{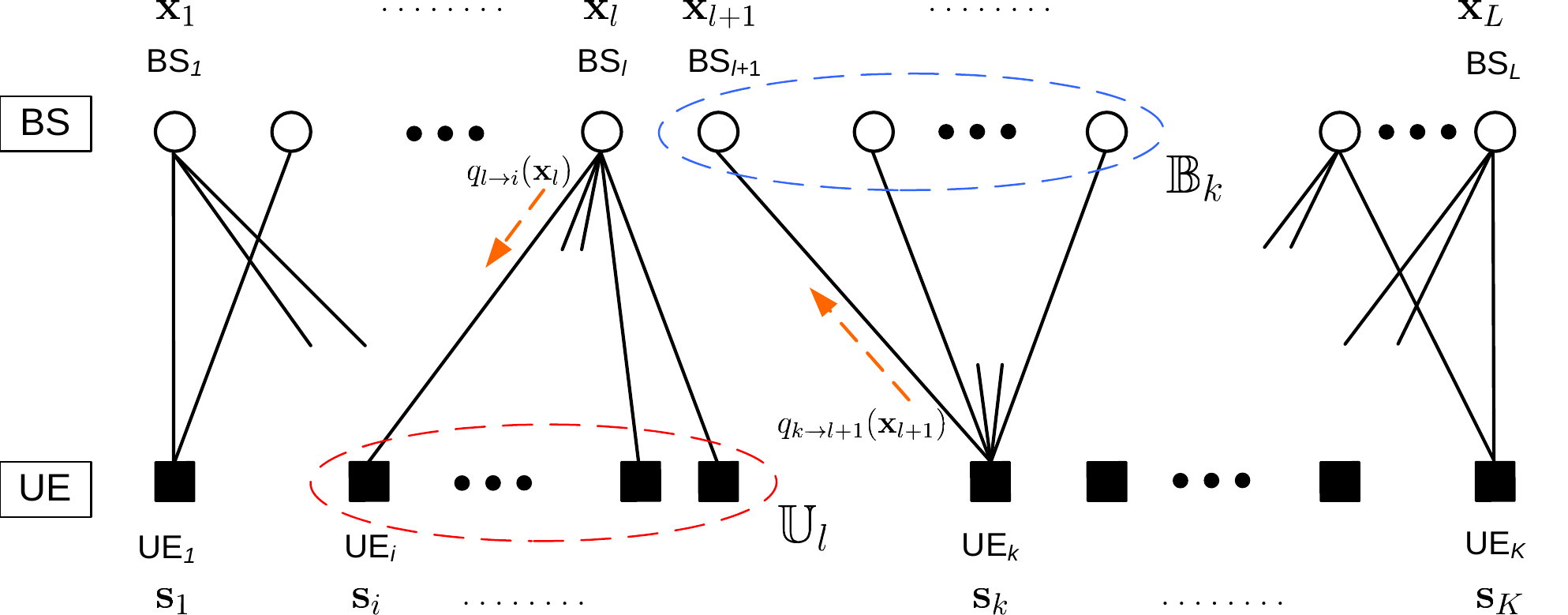} }%
\caption{A downlink model with BS cooperation.}\label{fig:MIMONetwork}
\end{center}
\end{figure}

As shown in Figure \ref{fig:MIMONetwork}, we consider a large-scale MIMO broadcast system where $L$ interconnected multi-antenna BSs, labeled as ${\sf BS}_1,\dots,{\sf BS}_L$, simultaneously
send information to $K$ users, labeled as ${\sf UE}_1,\dots,{\sf UE}_K$. In the system, ${\sf UE}_k$ is equipped with $M_k$ antennas while ${\sf BS}_l$ is equipped with $N_l$ antennas. Let $M
\triangleq \sum_{k=1}^{K} M_k$ and $N \triangleq \sum_{l=1}^{L} N_l$. The received signals at all the UEs can be expressed in a vector form as $\qy = [\qy_1^T,\ldots,\qy_K^T]^T \in \bbC^N$,
which is modeled as
\begin{equation} \label{eq:defSystem}
    \qy = \left[
    \begin{array}{ccc}
     \qH_{1,1} & \cdots & \qH_{1,L} \\
     \vdots & \ddots & \vdots \\
     \qH_{K,1} & \cdots & \qH_{K,L}
    \end{array}
    \right]
    \left[
    \begin{array}{c}
     \qx_{1}  \\
     \vdots \\
     \qx_{L}
    \end{array}
    \right]
     + \qz \triangleq \qH \qx + \qz,
\end{equation}
where $\qx_l$ denotes the transmitted signal from BS $l$, $\qH_{k,l}\in \bbC^{M_k \times N_l}$ represents the channel matrix from BS $l$ to UE $k$, and $\qz$ is the complex Gaussian noise vector
with zero mean and the covariance matrix $\sigma^2 \qI_M$. In (\ref{eq:defSystem}), we have defined $\qx \in \bbC^{N}$ as the vector of the transmitted signal and $\qH \in \bbC^{M \times N}$ as
the overall downlink channel matrix. Although (\ref{eq:defSystem}) appears to look like an $M\times N$ MIMO system, this is fundamentally different from a point-to-point MIMO channel. To see the
differences, we emphasize the following two features.

First, note that $\qH$ may have many zero block matrices because one UE is only able to receive signals from local BSs. The characteristic can be easily described via a graphical model as shown
in Figure \ref{fig:MIMONetwork}. For ease of expression, let $\bbU_l \subseteq \{ 1,2,\ldots,K\}$ comprise the set of user indices such that ${\sf BS}_l$ has some inference on these UEs; i.e.,
$\qH_{i,l} \neq \qzero$ for $i \in \bbU_l$. Similarly, let $\bbB_k \subseteq \{ 1,2,\ldots,L \}$ be a set of BS indices such that these BSs have some inference on ${\sf UE}_k$; i.e., $\qH_{k,j}
\neq \qzero$ for $j \in \bbB_k$. The local coupling is an important feature that communication should only be required between a subset of BSs rather than among all BSs.

Secondly, since both BS and UE are equipped with multiple antennas, the spatial correlation of the MIMO channel for each link between a BS and a UE should be considered. In this paper, we employ
the Kronecker model to characterize the spatial correlation of the MIMO channel for each link so that the correlation at a BS and a UE is modeled separately \cite{Shiu-00TCOM}. Specifically, the
channel from ${\sf BS}_l$ to ${\sf UE}_k$, $\qH_{k,l} \in{\mathbb C}^{M_k \times N_l}$, can be written as
\begin{equation}\label{eq:Spatial_Cov}
\qH_{k,l} = \qR_{k,l}^\frac{1}{2} \qW_{k,l} \qT_{k,l}^\frac{1}{2},
\end{equation}
where $\qR_{k,l} \in\mathbb{C}^{M_k \times M_k}$ and $\qT_{k,l} \in\mathbb{C}^{N_l \times N_l}$ are deterministic nonnegative definite matrices, which characterize the spatial correlation of the
received signals across the antenna elements of ${\sf UE}_k$ and that of the transmitted signals across the antenna elements of ${\sf BS}_l$ respectively, and $\qW_{k,l} \triangleq
[\frac{1}{\sqrt{N_l}} W_{ij}^{(k,l)}] \in\mathbb{C}^{M_k \times N_l}$ consists of the random components of the channel in which the elements $\{W_{ij}^{(l,k)}\}_{1\leq i \leq M_k; 1 \leq j \leq
N_l}$ are i.i.d.~complex Gaussian random variables with zero mean and unit variance. To get a proper definition on the channel gain of each link pair, we consider the power of the channel
\begin{equation}
\Ex \left\{\tr\left(\qH_{k,l}\qH_{k,l}^H\right)\right\} = \frac{1}{M_k} {\tr\left(\qR_k\right)} {\tr\left(\qT_k\right)} .
\end{equation}
If we assume that $\qR_{k,l}$ and $\qT_{k,l}$ are normalized such that ${\tr(\qR_{k,l})} = \varrho_{k,l} N_l$ and ${\tr(\qT_{k,l})} = M_k$, then $\varrho_{k,l}$ can be used as an indicator for
the link gain between ${\sf BS}_l$ and ${\sf UE}_k$.\footnote{ Indeed, the link gain can be included in either $\qR_{k,l}$ or $\qT_{k,l}$.}

In the broadcast system (\ref{eq:defSystem}), linear precoding, referred to as RZFBF, is used to project the data symbols onto a subspace using the $N$ transmit antennas. Let $\qs =[\qs_1^T,
\ldots, \qs_K^T]^T$ be the vector of data symbols, where $\qs_k$ corresponds to the data symbols intended for ${\sf UE}_k$. In RZFBF, the signal vector  transmitted by the BSs, denoted by $\qx$,
is given by \cite{Joham-02ISSSTA,Peel-05Tcom}
\begin{equation} \label{eq:Wprecoder}
    \qx = \alpha \qH^H (\qH\qH^H + \beta \qI_M)^{-1} \qs,
\end{equation}
where $\alpha$ is the normalization parameter to ensure that the transmit power constraint is met, i.e., $\Ex\{ \| \qx_l \|_2 \} \leq P_l$ for $l=1,\ldots,L$. Note that RZFBF can be regarded as
a generalization of other beamformers by adjusting the regularization parameter $\beta$. For instance, if $\beta=0$, it reduces to ZFBF whereas if $\beta\rightarrow\infty$, it will give the
matched-filter beamforming. Several other beamformers can also have a RZFBF structure by designing the regularization parameter appropriately \cite{Zak-12IT}. However, to obtain
(\ref{eq:Wprecoder}), all the BSs must cooperate to jointly process the data symbols from all the users in the network, requiring {\em global} CSI. If $N$ and $M$ are very large (which they
should in order to benefit from the gains of MU-MIMO and BS cooperation), the centralized approach will become prohibitively complex. Therefore, in the next section, we propose message passing
algorithms that can realize RZFBF in a distributed manner.

\section*{\sc III. A Bayesian Approach to Distributed RZFBF}
Now, we are concerning with the problem of distributing the computation of (\ref{eq:Wprecoder}) among the BSs. Toward this end, we first use the \emph{virtual} model concept of \cite{Ng-08IT},
which recasts the RZFBF optimization problem into an estimation problem which is described as follows.

Consider the virtual model
\begin{equation} \label{eq:auxChannel}
    \qs = \qH \qx + \tilde{\qz},
\end{equation}
where $\tilde{\qz} \in \mathbb{C}^{M}$ is the Gaussian random vector with zero mean and the covariance matrix $\beta \qI_M$. Recall that $\qs$ is the data symbol vector for all the users and
$\qx$ is the signal transmitted by the BSs. The virtual model implies that the transmitted signal $\qx$ goes through the channel $\qH$ and is then observed by $\qs - \tilde{\qz}$ at the UE
sides. What is important here is that the virtual model allows us to process the beamforming problem through a probabilistic inference approach. Specifically, we adopt a Bayesian approach.

The Bayes optimal way of estimating $\qx$ that minimizes the mean square error is given by \cite{Poor-94BOOK}
\begin{equation} \label{eq:estx}
    \hat{\qx} = \int \qx p(\qx|\qs) d\qx,
\end{equation}
where $p(\qx|\qs)$ is the posterior probability of $\qx$ given observation of $\qs$. Following Bayes theorem, we have
\begin{equation}
    p(\qx|\qs) = \frac{p(\qs|\qx) p(\qx)}{p(\qs)},
\end{equation}
where the conditional distribution of $\qs$ given $\qx$ under (\ref{eq:auxChannel}) is given by
\begin{equation} \label{eq:condPro}
    p(\qs|\qx) = \frac{1}{(\pi \beta)^N} e^{-\frac{1}{\beta}\sum_{k=1}^{K} \| \qs_k - \sum_{l \in \bbB_k} \qH_{k,l} \qx_l \|_2^2}.
\end{equation}
If we assume that $\qx$ is taken from the standard complex Gaussian random vector and its density is given by $p(\qx) = \frac{ e^{-\|\qx \|_2^2}}{\pi^{N}}$, then the posterior distribution
$p(\qx|\qs)$ admits an explicit expression as
\begin{equation} \label{eq:postPro}
    p(\qx|\qs) =  \frac{1}{\sfZ} ~e^{-\frac{1}{\beta}\sum_{k=1}^{K} \| \qs_k - \sum_{l \in \bbB_k} \qH_{k,l} \qx_l \|_2^2 - \sum_{l=1}^{L} \|\qx_l \|_2^2 },
\end{equation}
Henceforth, we shall use $\sfZ$ to denote a \emph{universal} normalization factor whose value may vary from one appearance to another. Plugging (\ref{eq:postPro}) into (\ref{eq:estx}) and
applying the Gaussian integral (Lemma \ref{Lemma 1} in Appendix B), one can get that the solution of (\ref{eq:estx}) is exactly identical to the form of (\ref{eq:Wprecoder}) without the power
normalization parameter. Next, we shall use an approach called BP (belief-propagation) for computing (\ref{eq:estx}).

\subsection*{A. BP-RZFBF}
We begin by applying the standard BP algorithm \cite{Kschischang-01IT} to perform (\ref{eq:estx}). As a matter of fact, BP can be regarded as a graphical method to estimate the marginal
distributions of the distribution $p(\qx|\qs)$ with respect to the variables $\qx_l$. To this end, we reformulate the problem as a bipartite graph called the factor graph. The corresponding
factor graph is depicted in Figure \ref{fig:MIMONetwork} where a circle represents a variable node associated with the transmit beamforming vector; i.e., $\qx_l$ for ${\sf BS}_l$, whereas a
square indicates a factor node associated with the sub-constraint function; i.e., $\|\qs_k- \sum_{l \in \bbB_k}\qH_{k,l}\qx_l\|_2^2 $ for ${\sf UE}_k$. There is an edge between a variable node
$l$ and a function node $k$ if and only if $\qH_{k,l} \neq\qzero$.

To estimate the marginal distributions, BP performs a set of message passing equations that go from factor nodes to variable nodes (i.e., ${k \rightarrow l}$) and from variable nodes to factor
nodes (i.e., ${l \rightarrow k}$) as is illustrated in Figure \ref{fig:MIMONetwork}. The message $q_{k \rightarrow l}$ from the factor node $k$ to the variable node $l$ is the marginal
probability of the variable $\qx_l$ when only the sub-constraint $k$ is present. On the other hand, the message $q_{l \rightarrow k}$ from the variable node $l$ to the factor node $k$ is the
marginal probability of the variables $\qx_l$ in the absence of the sub-constraint $k$.

Specifically, in order to estimate these marginal distributions $p(\qx_l|\qs)$ with BP algorithm, $2KL$ messages for the probability distributions of the variables $\qx_l$ are constructed in the
following way \cite{Kschischang-01IT}:
\begin{subequations} \label{eq:Standard_BP}
\begin{align}
    q_{k \rightarrow l}^{(t)}(\qx_l) &= \frac{1}{Z_{k \rightarrow l}} \int{ \prod_{j \in \bbB_k\backslash l} d\qx_j q_{j\rightarrow k}^{(t-1)}(\qx_j)
    e^{-\frac{1}{\beta} \| \qs_k - \sum_{j \in \bbB_k\backslash l} \qH_{k,j} \qx_j - \qH_{k,l} \qx_l \|_2^2 } }, \label{eq:qkl} \\
    q_{l \rightarrow k}^{(t-1)}(\qx_l) &= \frac{1}{Z_{l \rightarrow k}} p(\qx_l) \prod_{i \in \bbU_l\backslash k} q_{i \rightarrow l}^{(t-1)}(\qx_l), \label{eq:qlk}
\end{align}
\end{subequations}
where $t = 1,2, \ldots$ represents the iteration index and $Z^{k \rightarrow l}$ and $Z^{l \rightarrow k}$ are the normalization factors ensuring that $\int d\qx_l q_{k \rightarrow
l}^{(t)}(\qx_l) = \int d\qx_l q_{l \rightarrow k}^{(t-1)}(\qx_l) = 1$. At the termination of the message passing algorithm, say at iteration $T$, the final estimate of $\qx_l$ is given by
$\hat{\qx}_l = \int \qx_l q_{l}^{(T)}(\qx_l) d\qx_l$ where $q_{l}^{(T)}(\qx_l) \propto \prod_{k=1}^K q_{k \rightarrow l}^{(T)}(\qx_l)$. The RZFBF solution can thus be realized in a distributed
manner via the message passing procedures. However, the messages are {\em density functions} which are usually too complex to be exchanged and will cost a huge burden in the backhaul in our
application of BS cooperation.

To overcome this, the message can be approximated by Gaussian and parameterized by the mean and covariance. Instead of passing the density functions, we thus have the mean and covariance as the
messages:
\begin{align}
    \qx_{l \rightarrow k}^{(t)} &= \left\langle \qx_l \right\rangle_{q_{l \rightarrow k}^{(t)}}, \label{eq:def_xlk}\\
    \qV_{l \rightarrow k}^{(t)} &= \left\langle (\qx_l-\qx_{l \rightarrow k}^{(t)})(\qx_l-\qx_{l \rightarrow k}^{(t)})^H \right\rangle_{q_{l \rightarrow k}^{(t)}}, \label{eq:def_Vlk}
\end{align}
where $\ang{ f(\qx_l) }_{q_{l \rightarrow k}}$ denotes the average or expectation of a function $f(\qx_l)$ over the random vector $\qx_l$ with distribution $q_{l \rightarrow k}(\qx_l)$.
Mathematically, that is
\begin{equation*}
    \ang{ f(\qx_l) }_{q_{l \rightarrow k}} \triangleq \int  f(\qx_l) q_{l \rightarrow k}(\qx_l) d\qx_l.
\end{equation*}
The Gaussian approximation method was introduced in \cite{Rangan-10ArXiv,Guo-06ITW} when the message is scalar and the concerned matrix is sparse. In our case, we
follow the techniques in \cite{Krzakala-12JSM} by considering that the block matrix $\qH_{k,l}$ scales as $O(1/\sqrt{N_l})$. As a consequence, we can approximate
$q_{k \rightarrow l}(\qx_l)^{(t)}$ by
\begin{equation} \label{eq:appqkl}
    \tilde{q}_{k \rightarrow l}^{(t)}(\qx_l) \propto e^{- \left(\qx_l^H \qE_{k \rightarrow l}^{(t)} \qx_l  - (\qF_{k \rightarrow l}^{(t)})^H \qx_l - \qx_l^H\qF_{k \rightarrow l}^{(t)} \right) },
\end{equation}
where
\begin{align}
    \qE_{k \rightarrow l}^{(t)} &= \qH_{k,l}^H\left( \sum_{j \in \bbB_k\backslash l} \qH_{k,j} \qV_{j \rightarrow k}^{(t-1)} \qH_{k,j}^H + \beta \qI_{M_k} \right)^{-1}\qH_{k,l}, \label{eq:Edef} \\
    \qF_{k \rightarrow l}^{(t)} &= \qH_{k,l}^H \left( \sum_{j \in \bbB_k\backslash l} \qH_{k,j} \qV_{j \rightarrow k}^{(t-1)} \qH_{k,j}^H + \beta \qI_{M_k} \right)^{-1}
    \left( \qs_k -  \sum_{j \in \bbB_k\backslash l} \qH_{k,j} \qx_{j \rightarrow k}^{(t)} \right). \label{eq:Fdef}
\end{align}
Notice that $\qx_{j \rightarrow k}$ and $\qV_{j \rightarrow k}$ in (\ref{eq:Edef})--(\ref{eq:Fdef}) are functions of $q_{j \rightarrow k}$ which is also altered due to the approximation
$\tilde{q}_{k \rightarrow l}$. To make the connection, from (\ref{eq:appqkl}) and (\ref{eq:qlk}), we have
\begin{equation} \label{eq:tqlk}
    \tilde{q}_{l \rightarrow k}^{(t)}(\qx_l) \propto
    p(\qx_l) e^{- \sum_{i \in \bbU_l\backslash k}\left(\qx_l^H \qE_{i \rightarrow l}^{(t)} \qx_l  - (\qF_{i \rightarrow l}^{(t)})^H \qx_l - \qx_l^H \qF_{i \rightarrow l}^{(t)}  \right) }.
\end{equation}
Henceforth, we will replace $\qx_{l \rightarrow k}^{(t)}$ and $\qV_{l \rightarrow k}^{(t)}$ in (\ref{eq:Edef})--(\ref{eq:Fdef}) by $\tilde{\qx}_{l \rightarrow k}^{(t)}$ and $\tilde{\qV}_{l
\rightarrow k}^{(t)}$ which are, respectively, the mean and covariance over the probability distribution $\tilde{q}_{l \rightarrow k}^{(t)}$. Also, in the sequel, we will no longer use the
probability distribution $q_{l \rightarrow k}^{(t)}$. However, for notational convenience, we will abuse our notation slightly and still use $\qx_{l \rightarrow k}^{(t)}$ and $\qV_{l \rightarrow
k}^{(t)}$ to denote those mean and covariance over the probability distribution $\tilde{q}_{l \rightarrow k}^{(t)}$.

Recall that $\qx_l$ is taken from the standard complex Gaussian random vector. By applying the Gaussian integral (Lemma \ref{Lemma 1} in Appendix B), $\qx_{l\rightarrow k}$ and $\qV_{l
\rightarrow k}$ with the distribution $\tilde{q}_{l \rightarrow k}(\qx_l)$ in (\ref{eq:tqlk}) can be computed analytically. These lead to the following closed form of the BP update:
\begin{subequations} \label{eq:BPupdate}
\begin{align}
    \qx_{l \rightarrow k}^{(t)} &=  \left(\qwbE_{l\backslash k}^{(t)} + \qI_{N_l} \right)^{-1} \qwbF_{l\backslash k}^{(t)}, \\
    \qV_{l \rightarrow k}^{(t)} &= \left(\qwbE_{l\backslash k}^{(t)} + \qI_{N_l} \right)^{-1},
\end{align}
\end{subequations}
where $\qwbE_{l\backslash k}^{(t)} \triangleq \sum_{i \in \bbU_l\backslash k}\qE_{i \rightarrow l}^{(t)} $ and $\qwbF_{l\backslash k}^{(t)} \triangleq \sum_{i \in\bbU_l\backslash
k}\qF_{i\rightarrow l}^{(t)} $. The number of messages is still $2 K L$. However, the message update here is only on the mean and covariance rather than the functional update in
(\ref{eq:Standard_BP}). At the termination of the BP, the final estimation of $\qx_{l}$ is given by
\begin{equation}
    \qx_{l}^{(t)} = \left( \qwbE_{l}^{(t)} + \qI_{N_l} \right)^{-1} \qwbF_{l}^{(t)},
\end{equation}
where $\qwbE_{l}^{(t)} \triangleq \sum_{i \in \bbU_l}\qE_{i \rightarrow l}^{(t)} $ and $\qwbF_{l}^{(t)} \triangleq \sum_{i \in \bbU_l}\qF_{i \rightarrow l}^{(t)} $. We refer to this algorithm as
BP-RZFBF although a variant of BP is adopted here. In some applications, the regularization parameter $\beta$ varies for different UEs. In these cases, we only have to simply replace $\beta$
with $\beta_k$ in (\ref{eq:Edef})--(\ref{eq:Fdef}).

BP-RZFBF is a generalization of \cite[(27)--(28)]{Ng-08IT} in which $\qH_{k,l}$'s are scalars (real numbers). Clearly, this generalization can be applied to a wide range of scenarios with
complex-valued systems. Additionally, it performs block matrix computations resulting in a natural partition of BSs.

\subsection*{B. AMP-RZFBF}
In BP-RZFBF, each BS has to send separate messages with respect to $k$; i.e., $\qx_{l \rightarrow k}^{(t)}$ and $\qV_{l \rightarrow k}^{(t)}~\forall k$. We can reduce the messaging overhead to
$2 (K + L)$. To do so, we note that the messages $\qx_{l \rightarrow k}^{(t)}$ and $\qV_{l\rightarrow k}^{(t)}$ are functions of $\qwbE_{l\backslash k}^{(t)}$ and $\qwbF_{l\backslash k}^{(t)}$
which are nearly independent of $k$. However, one must keep all the correction terms that are linear in $\qH_{l,k}$. This methodology was first introduced in compressed sensing applications in
\cite{Donoho-09PNAS} and is referred to as AMP. Using AMP in the BF-RZFBF problem, we have developed the AMP-RZFBF algorithm in Algorithm \ref{ago:agoAMP}. For readability, we give the detailed
derivation in Appendix A.

\begin{algorithm}[h]\label{ago:agoAMP}  \footnotesize
  \caption{AMP-RZFBF}
  \KwIn{ Data symbols $\qs_k$ for $k=1,\ldots,K$, channel matrices $\qH_{k,l}$ for $k=1,\ldots,K$ and $l=1,\ldots,L$. }
  \KwOut{Return the RZFBF $\qx_l$ for $l=1,\ldots,L$. }
  \Begin{
  Select $\qx_l^{(0)} = \qzero$, $\qV_l^{(0)} = \qI_{N_l}$, and $\qnu_k^{(0)} = \qs_k$ for $k=1,\ldots,K$ and $l=1,\ldots,L$\;
  $t \Longleftarrow 1$ \\
  \Repeat{Predefined number of iterations is met}{
  $\qOmega_k^{(t)} = \sum_{l \in \bbB_k} \qH_{k,l} \qV_l^{(t-1)} \qH_{k,l}^H \label{eq:TBSeqs_a}$ \;
  $\qnu_k^{(t)} = \qs_k - \sum_{l \in \bbB_k} \qH_{k,l}\qx_l^{(t-1)} +  \left(\qOmega_k^{(t-1)}+\beta \qI_{M_k}\right)^{-1}\qOmega_k^{(t)} \qnu_k^{(t-1)} \label{eq:TBSeqs_b}$ \;
  $\qSigma_l^{(t)} = \sum_{k \in \bbU_l} \qH_{k,l}^H\left(\qOmega_k^{(t)}+\beta\qI_{M_k}\right)^{-1}\qH_{k,l} \label{eq:TBSeqs_c} $ \;
  $\qmu_l^{(t)} = \qx_l^{(t-1)} + \left(\qSigma_l^{(t)}\right)^{-1} \left[ \sum_{k \in \bbU_l} \qH_{k,l}^H \left(\qOmega_k^{(t)}+\beta\qI_{M_k}\right)^{-1} \qnu_k^{(t)} \right] \label{eq:TBSeqs_d}$ \;
  $\qx_l^{(t)} = \left(\qSigma_l^{(t)} + \qI_{N_l} \right)^{-1} \qSigma_l^{(t)} \qmu_l^{(t)} \label{eq:TBSeqs_e}$ \;
  $\qV_l^{(t)} = \left(\qSigma_l^{(t)} + \qI_{N_l} \right)^{-1} \label{eq:TBSeqs_f}$ \;
  $t \Longleftarrow t+1$
  }
  }
\end{algorithm}

Now, we turn our attention to realizing AMP-RZFBF for the cooperative system. In general, each iteration requires a broadcast and gathering operation. We assume that each BS has local data
information and CSI; e.g., only $\{ \qs_k, \qH_{l,k} \}$ for $k \in \bbU_l$ are known at ${\sf BS}_l$. The first two steps of AMP-RZFBF consist of performing $\qOmega_k^{(t)}$ and $\qnu_k^{(t)}$
updates at ${\sf BS}_l$. Notice that for ${\sf BS}_l$, $\qOmega_k^{(t)}$ and $\qnu_k^{(t)}$ updates are only for indices $k \in \bbU_l$ which correspond to the user indices within its reception
range. In order to update $\qOmega_k^{(t)}$ and $\qnu_k^{(t)}$, ${\sf BS}_l$ must gather $\qH_{k,l} \qV_l^{(t-1)} \qH_{k,l}^H$ and $\qH_{k,l}\qx_l^{(t-1)}$ from the set of its neighboring BSs
$\bbB_k$. After getting $\qOmega_k^{(t)}$ and $\qnu_k^{(t)}$, ${\sf BS}_l$ is able to compute $(\qSigma_l^{(t)}, \qmu_l^{(t)})$ and then update $(\qx_l^{(t)}, \qV_l^{(t)})$ subsequently. Once
$(\qx_l^{(t)}, \qV_l^{(t)})$ are computed, ${\sf BS}_l$ will broadcast $\qH_{k,l} \qV_l^{(t)} \qH_{k,l}^H$ and $\qH_{k,l}\qx_l^{(t)}$ to its neighboring BSs. The algorithm continues to repeat
the procedures above until it reaches a predefined number of iterations.

In AMP-RZFBF, the computation of $\qSigma_l^{(t)}$ and $\qV_k^{(t)}$ involves several matrix inversions for every channel realization. These demand high computational cost and rapid information
exchange between the BSs. To remedy this, we propose to infer these parameters based on CCoI which varies much slower than CSI.

\subsection*{C. CCoI-aided AMP-RZFBF}
Starting from the initial condition, we approximate $\qOmega_l^{(1)}$ by its average with respect to different realization of the measurement matrix $\qH_{k,l}$:
\begin{align} \label{eq:U_teq1}
    \qOmega_k^{(1)} \approx \Ex\left\{ \qOmega_k^{(1)} \right\}
    &= \sum_{l \in \bbB_k}  \frac{1}{N_l}\tr(\qT_{k,l})  \qR_{k,l},
\end{align}
where the equality follows from Lemma \ref{Lemma 2} in Appendix B. The approximation is benefited by the self-averaging property in statistical physics; that is, a quantity per degree of freedom
has small deviations from its mean. In fact, using techniques from random matrix theory, e.g., \cite{Bai-10}, one can show that as $N_l \rightarrow \infty$, $\qOmega_k^{(1)} \rightarrow \Ex\{
\qOmega_k^{(1)} \}$ almost surely. We find it useful to denote $\tilde{\varsigma}_{k,l}^{(1)} = \frac{1}{N_l} \tr(\qT_{k,l})$ and define
\begin{equation} \label{eq:defR}
\qwbR_k^{(t)} \triangleq \sum_{l \in \bbB_k}{ \tilde{\varsigma}_{k,l}^{(t)} \qR_{k,l} }.
\end{equation}
Applying the similar argument to $\qSigma_l^{(t)}$ in Line \ref{eq:TBSeqs_c} of Algorithm \ref{ago:agoAMP} for $t=1$, we have
\begin{equation} \label{eq:Sigma_te1}
    \qSigma_l^{(t)} \approx \sum_{k \in \bbU_l} \frac{1}{N_l}\tr{\left( \qR_{k,l}\left(\qwbR_k^{(t)} + \beta\qI_{M_k}\right)^{-1}\right)}\qT_{k,l}.
\end{equation}
Again, for ease of expression, we also define
\begin{equation} \label{eq:defVarsigma_t}
    \varsigma_{k,l}^{(t)} \triangleq \frac{1}{N_l} \tr{\Big( \qR_{k,l} (\qwbR_k^{(t)} + \beta \qI_{M_k} )^{-1} \Big)},
\end{equation}
and
\begin{equation}
    \qwbT_l^{(t)} \triangleq \sum_{k \in \bbU_l}{ \varsigma_{k,l}^{(t)}  \qT_{k,l}}.
\end{equation}
Substituting the above definitions, (\ref{eq:Sigma_te1}) is then expressed as
\begin{equation} \label{eq:Sigma_t}
    \qSigma_l^{(t)} \approx \qwbT_l^{(t)} .
\end{equation}
Now, $\qx_l^{(t)}$ and $\qV_l^{(t)}$ can be calculated as those in Lines \ref{eq:TBSeqs_e}--\ref{eq:TBSeqs_f} of Algorithm \ref{ago:agoAMP} but $\qV_l^{(t)} = \left(\qSigma_l^{(t)} + \qI_{N_l}
\right)^{-1}$ is approximated by
\begin{equation} \label{eq:appV_t}
    \qV_l^{(t)} \approx  \left( \qwbT_l^{(t)} + \qI_{N_l} \right)^{-1}.
\end{equation}
Note that when $t=1$, $\qOmega_l^{(t)}$ is given by (\ref{eq:U_teq1}). Let us go ahead on the next round of iteration to get a general expression for $\qOmega_l^{(t)}$ for general $t$. Following
the similar argument as that used in (\ref{eq:U_teq1}), we have
\begin{equation} \label{eq:U_t1}
    \qOmega_k^{(t)} \approx \Ex\left\{ \qOmega_k^{(t)} \right\}
    = \sum_{l \in \bbB_k}  \frac{1}{N_l}\tr\left( \qT_{k,l} \left(\qwbT_{k}^{(t-1)}+\qI_{N_l} \right)^{-1} \right) \qR_{k,l}.
\end{equation}
Define
\begin{equation} \label{eq:deftVarsigma_t}
    \tilde{\varsigma}_{k,l}^{(t)} \triangleq \frac{1}{N_l} \tr{\Big( \qT_{k,l} \left(\qwbT_l^{(t-1)} + \qI_{N_l} \right)^{-1} \Big)}.
\end{equation}
Then (\ref{eq:U_t1}) becomes
\begin{equation} \label{eq:U_t2}
    \qOmega_l^{(t)} \approx     \sum_{k \in \bbU_l} \tilde{\varsigma}_{k,l}^{(t)} \qR_{k,l}.
\end{equation}

Recall that the updates of $\qOmega_l^{(t)}$, $\qSigma_l^{(t)}$, and $\qV_l^{(t)}$ in Lines \ref{eq:TBSeqs_a}, \ref{eq:TBSeqs_c}, and \ref{eq:TBSeqs_f} of
Algorithm \ref{ago:agoAMP}, respectively, involve the channel realizations $\{ \qH_{k,l} \}$. These computations are replaced by (\ref{eq:U_t2}),
(\ref{eq:Sigma_t}), and (\ref{eq:appV_t}), where only the CCoI is required. Therefore, Algorithm \ref{ago:agoAMP} together with these replacements lead to the
simpler iteration forms. The algorithmic description of this CCoI-aided AMP-RZFBF is summarized in Algorithm \ref{ago:agoCCIAMP}.

\begin{algorithm}[h]\label{ago:agoCCIAMP}  \footnotesize
  \caption{CCI-aided AMP-RZFBF}
  \KwIn{ Data symbols $\qs_k$ for $k=1,\ldots,K$, channel matrices $\qH_{k,l}$ for $k=1,\ldots,K$ and $l=1,\ldots,L$,
  and CCI $\{ \qT_{k,l}, \qR_{k,l} \}$  for $k=1,\ldots,K$ and $l=1,\ldots,L$.}
  \KwOut{Return the RZFBF $\qx_l$ for $l=1,\ldots,L$ }

  \Begin{
  Select $\qx_l^{(0)} = \qzero$, $\qnu_k^{(0)} = \qs_k$, $\qwbR_k^{(0)} = \qzero$, and $\qwbT_l^{(0)} = \qzero$ for $k=1,\ldots,K$ and $l=1,\ldots,L$\;
  $t \Longleftarrow 1$ \\
  \Repeat{Predefined number of iterations is met}{
  $\tilde{\varsigma}_{k,l}^{(t)} = \frac{1}{N_l} \tr{\Big( \qT_{k,l} \left(\qwbT_l^{(t-1)} + \qI_{N_l} \right)^{-1} \Big)}$ \;
  $\qwbR_k^{(t)} = \sum_{l \in \bbB_k}{ \tilde{\varsigma}_{k,l}^{(t)} \qR_{k,l} }$ \;
  $\varsigma_{k,l}^{(t)} = \frac{1}{N_l} \tr{\Big( \qR_{k,l} \left(\qwbR_k^{(t)} + \beta \qI_{M_k} \right)^{-1} \Big)}$ \;
  $\qwbT_l^{(t)} = \sum_{k \in \bbU_l}{ \varsigma_{k,l}^{(t)} \qT_{k,l}}$ \;
  $\qA_l^{(t)} = \qwbT_l^{(t)}\left(\qwbT_l^{(t)} + \qI_{N_l}\right)^{-1}$ \;
  $\qB_k^{(t)} = \qwbR_k^{(t)} \left(\qwbR_k^{(t-1)} + \beta \qI_{M_k} \right)^{-1}$ \;
  $\qnu_k^{(t)} =  \qs_k - \sum_{l \in \bbB_k}{ \qH_{k,l} \qx_l^{(t-1)} } +  \qB_k^{(t)} \qnu_k^{(t-1)}$ \;
  $\qx_l^{(t)} = \qA_l^{(t)} \left[ \qx_l^{(t-1)}
    + \left(\qwbT_l^{(t)}\right)^{-1} \left( \sum_{k \in \bbU_l}{  \qH_{k,l}^H \left(\qwbR_k^{(t)} + \beta \qI_{M_k} \right)^{-1} \qnu_{k}^{(t)} } \right) \right]$
  \;

  $t \Longleftarrow t+1$
  }
  }
\end{algorithm}

The realization of CCoI-aided AMP-RZFBF is similar to that of AMP-RZFBF but with much lower computational complexity and much less communication overhead. Firstly, notice that lines 5--10 of
Algorithm \ref{ago:agoCCIAMP} can be computed \emph{offline} and \emph{locally} regardless of the channel realizations $\{ \qH_{k,l} \}$, data symbols $\{ \qs_k \}$, and the outputs
$\{\qx_l^{(t)} \}$ of each iteration. Because CCoI can be considered static, the BSs compute and exchange these parameters at the time scale at which the CCoI changes rather than the
instantaneous channel realizations. This characteristics significantly reduces the computational complexity and the communication overhead. Secondly, the remaining two steps, lines 11--12 of
Algorithm \ref{ago:agoCCIAMP}, involve only linear matrix multiplications. The update of $\qnu_k^{(t)}$ and $\qx_l^{(t)}$ also requires a general broadcast and gathering operation. In
particular, to update $\qnu_k^{(t)}$, ${\sf BS}_l$ must gather $\qH_{k,l} \qx_l^{(t-1)}$ from the set of BSs $\bbB_k$ but it only updates $\qnu_k^{(t)}$ for $k \in \bbU_l$. After $\qx_l^{(t)}$
is computed, ${\sf BS}_l$ will broadcast $\qH_{k,l}\qx_l^{(t)}$ to its neighboring BSs. The algorithm continues to repeat the procedures above until it reaches a predefined number of iterations.

\section*{\sc IV. Simulation Results}
In this section, we compare the performance of different algorithms through simulations. The considered algorithms include all the message passing algorithms in Section III (i.e., BP-RZFBF,
AMP-RZFBF, and CCoI-aided AMP-RZFBF) and the ADMM approach in \cite[Section 8.3]{Boyd-11BOOK}. ADMM is the state-of-the-art optimization technique and has now been widely used in performing
distributed estimations.

Before proceeding, let us first take a look at the computational complexity of these algorithms. In BP-RZFBF, most of computational complexity lies in the matrix inversions in
(\ref{eq:Edef})--(\ref{eq:Fdef}) and (\ref{eq:BPupdate}). Moreover, we have to perform these matrix inversions for all the $ 2 K L$ messages. This gives a complexity of order $  KL O(M_k^3)$ for
each iteration. In AMP-RZFBF, the complexity also lies in the matrix inversions while the messaging overhead is reduced to $2 (K + L)$. Therefore, the complexity of AMP-RZFBF is of order $ (K+L)
O(M_k^3)$ for each iteration. The complexity of the ADMM approach is comparable to AMP-RZFBF. Finally, the complexity of CCoI-aided AMP-RZFBF is further reduced from AMP-RZFBF because the matrix
inversions are performed at the time scale at which the CCoI changes. Therefore, the computational complexity of CCoI-aided AMP-RZFBF is of order $ \left(\frac{K+L}{\tau}\right) O(M_k^3)$ for
each iteration where $\tau$ represents the time scale at which the CCoI changes. The value of $\tau$ could be very large because CCoI can be considered static. Consequently, CCoI-aided AMP-RZFBF
can be implemented in the most efficient way.

\begin{figure}
\begin{center}
\resizebox{5.0in}{!}{%
\includegraphics*{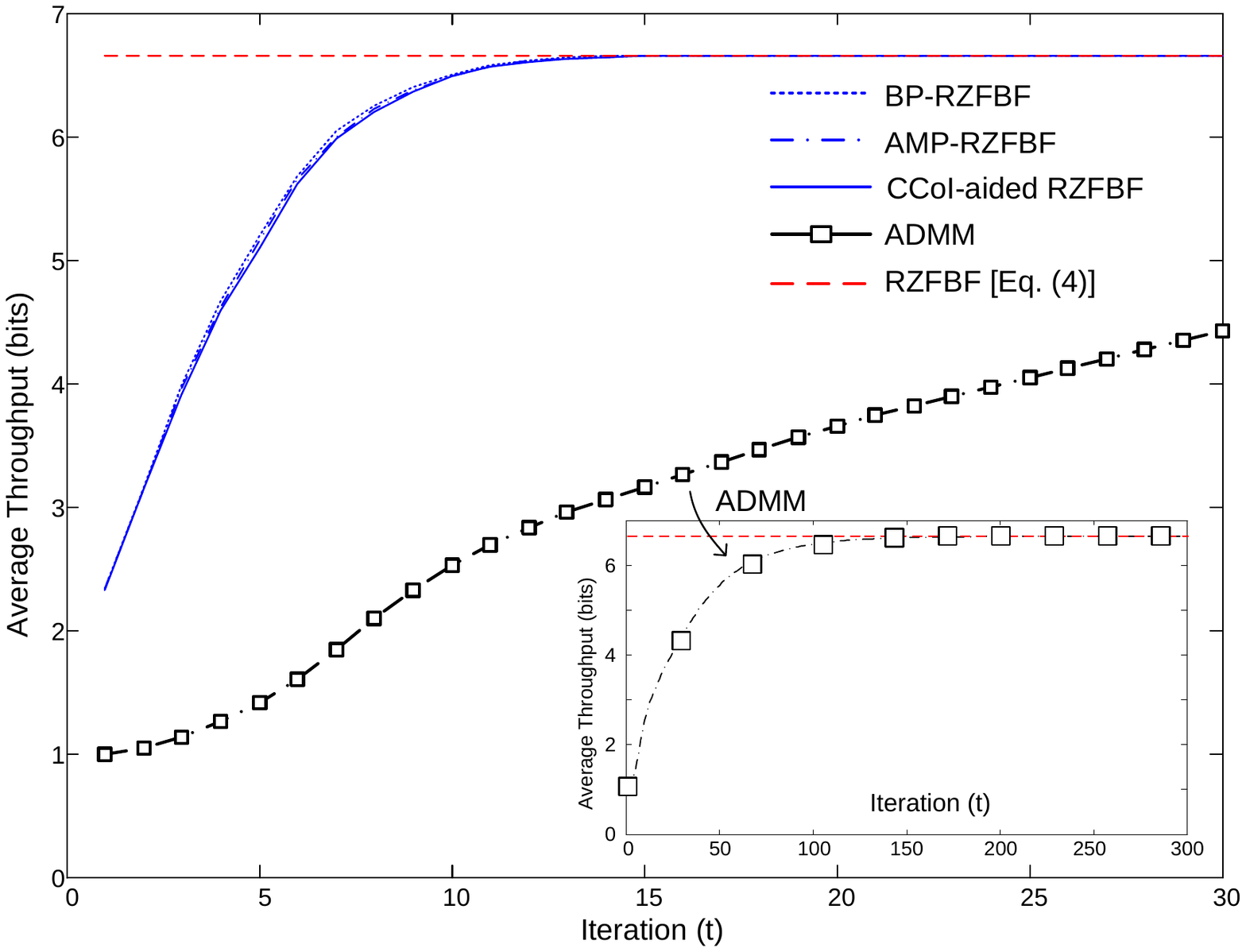} }%
\caption{Average throughput against the number of iterations for the message passing beamformers and global beamformer when $L=100$, $K=100$, $N_l =8$, $M_k = 4$, and $\beta = \sigma^2 = 10^{-2}$.}\label{fig:Fig1_L100K100M8N4}
\end{center}
\end{figure}

With the computational complexity in mind, our attention turns to their performances. We consider a cellular system with $100$ BSs and $100$ UEs in which each BS is equipped with $8$ transmit
antennas and each user has $4$ receive antennas, i.e., $L=100$, $K=100$, $N_l =8$, and $M_k = 4$. The propagation channel matrix between each BS and UE is characterized by
(\ref{eq:Spatial_Cov}), where the spatial correlations $\qR_{k,l}$'s and $\qT_{k,l}$'s are arbitrarily generated with elements being $[\qR_{k,l}]_{i,j} = \rho_{{\sf R}_{k,l}}^{|i-j|}$ and $
\qT_{k,l}]_{i,j} = \rho_{{\sf T}_{k,l}}^{|i-j|}$, respectively. Additionally, the link gain $\varrho_{k,l}$ is included in $\qR_{k,l}$ and is also uniformly and randomly generated. Figure
\ref{fig:Fig1_L100K100M8N4} illustrates the average throughput of the algorithms varies with the number of message transfers. The average throughput is calculated by $\frac{1}{M} \sum_{k=1}^{K}
\sum_{m=1}^{M_k} \log_2\left( 1+ \gamma_{m,k}^{(t)} \right)$ where $\gamma_{m,k}^{(t)} \triangleq \frac{| \qe_m^T\qs_{k} |^2}{ | \qe_m^T (\qH_k \qx^{(t)} - \qs_{k}) |^2+\sigma^2}$ and
$\qx^{(t)}$ is the vector of transmitted signals at the $t$-th iteration. Here, $\qH_k$ denotes $[\qH_{k,1} \cdots \qH_{k,L}]$ and $\qe_m$ has been defined in Notations. The results provided are
for a particular realization of the channel. It is natural that when the number of iterations increases, the average throughput increases and saturates eventually. Here, RZFBF in
(\ref{eq:Wprecoder}) serves as a benchmark for the optimal beamformer. From Figure \ref{fig:Fig1_L100K100M8N4}, it can be observed that the proposed message passing algorithms converge
significantly faster than the ADMM approach. The convergence rates of all the proposed message passing algorithms are very similar.

\begin{figure}
\begin{center}
\resizebox{5.0in}{!}{%
\includegraphics*{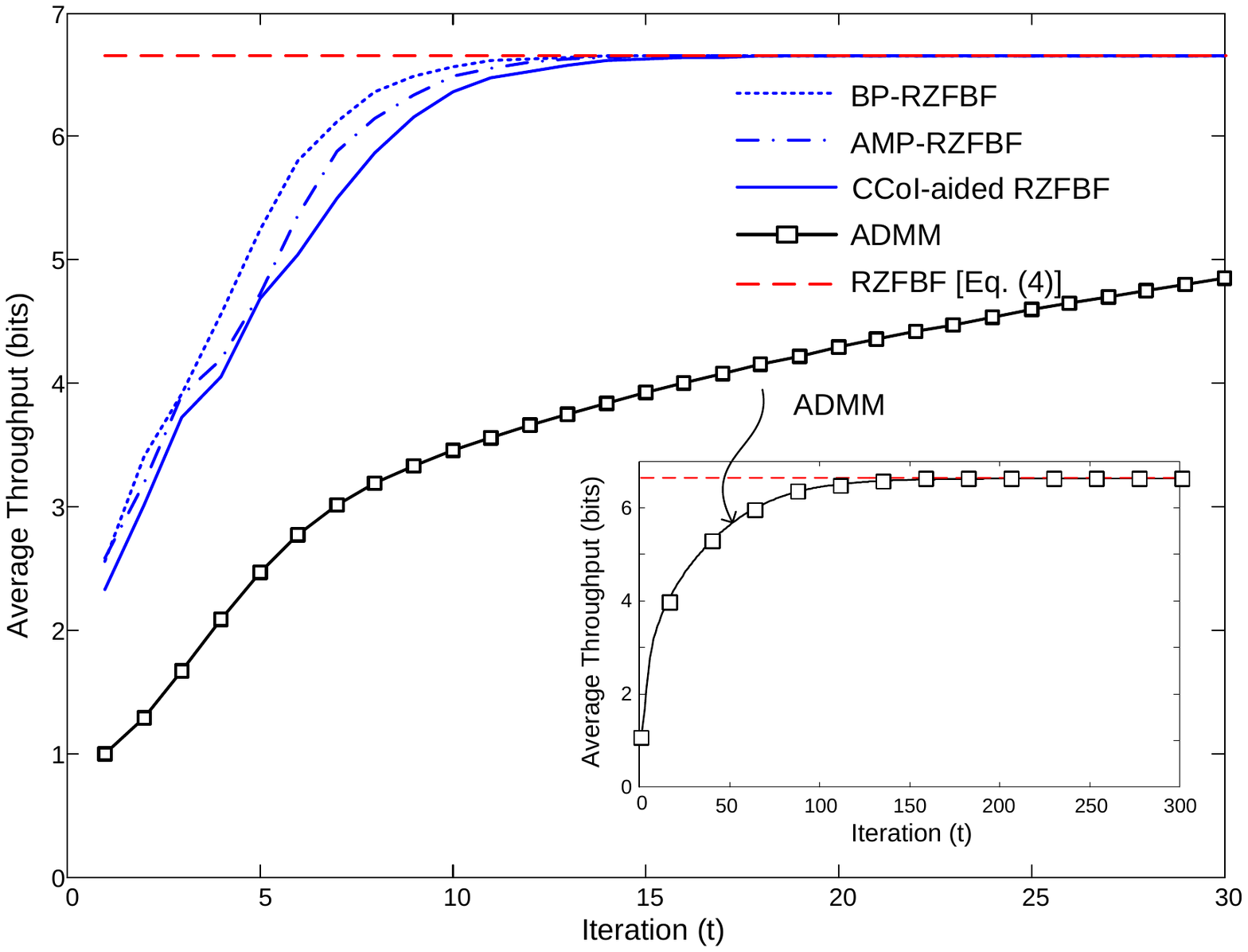} }%
\caption{Average throughput against the number of iterations for the message passing beamformers and global beamformer when $L=16$, $K=16$, $N_l = 4$, $M_k = 2$, and $\beta = \sigma^2 = 10^{-2}$.}\label{fig:Fig2_L10K10M4N2}
\end{center}
\end{figure}

Recall that AMP-RZFBF follows from BP-RZFBF but using the approximations that $\qwbE_{l\backslash k}$ and $\qwbF_{l\backslash k}$ are nearly independent of $k$. This approximation is expected to
be good if $K$ and $L$ are extremely large. Furthermore, the CCoI-aided AMP-RZFBF uses the large system approximation by assuming $N_l \rightarrow \infty$. Although the setting in Figure
\ref{fig:Fig1_L100K100M8N4} corresponds to a practical system dimension, it is intriguing to see their performances under a relatively small network; e.g., $L=16$, $K=16$, $N_l = 4$, and $M_k =
2$. Under the small network consideration, Figure \ref{fig:Fig2_L10K10M4N2} illustrates the convergence of the algorithms. Similar characteristics as in Figure \ref{fig:Fig1_L100K100M8N4} before
are observed. Additionally, comparing to BP-RZFBF, AMP-RZFBF and CCoI-aided AMP-RZFBF only slightly degrades the convergence rate. This result is quite different from several earlier designs
based on CCoI, e.g., \cite{Hoydis-11ISIT,Lakshminarayana-12SP}. Usually, when some calculations are approximated by the CCoI, an obvious degradation in performance would be observed but this is
not the case in our scheme.

\section*{\sc V. Conclusion}
Using Bayesian inference, this paper proposed several message passing algorithms for realizing RZFBF in cooperative-BS networks, namely, BP-RZFBF, AMP-RZFBF and CCoI-aided AMP-RZFBF. Results
showed that the proposed algorithms converge very fast to the exact RZFBF. Comparing to BP-RZFBF, both AMP-RZFBF and CCoI-aided AMP-RZFBF perform well with only very slight degradation in the
convergence rate, but greatly reducing the burden for information exchange between the BSs.

\section*{Appendix A: Derivation for AMP-RZFBF}
To derive AMP-RZFBF, we use a heuristic approximation which keeps all the terms that are linear in the matrix $\qH_{l,k}$ while neglecting the higher-order terms. The similar methodology was
used in \cite{Krzakala-12JSM} in the case of compressed sensing although some modifications are required to reflect the concerned case.

We start by noticing that $\qwbE_{l\backslash k}^{(t)} = \sum_{i \neq k}\qE_{i \rightarrow l}^{(t)} $ is the sum of $K$ terms each of order $1/N_l$ because $\qH_{l,k} $ scales as
$O(1/\sqrt{N_l})$. Therefore, it is natural to approximate $\qwbE_{l\backslash k}^{(t)}$ by $\qwbE_{l}^{(t)} = \sum_{i}\qE_{i \rightarrow l}^{(t)} $ which only depends on the index $l$ and not
on $k$. Similarly, it is natural to anticipate a similar approximation for $\qwbF_{l\backslash k}^{(t)}$. However, we must be careful to keep all correction terms of order $1/\sqrt{N_l}$. To
that end, we instead set
\begin{equation}
    \qwbF_{l\backslash k}^{(t)}= \qwbF_{l}^{(t)} - \qDelta \qwbF_{l \rightarrow k}^{(t)}.
\end{equation}
Recall from (\ref{eq:BPupdate}) that $\qx_{l \rightarrow k}^{(t)} = \left(\qwbE_{l\backslash k}^{(t)} + \qI_{N_l} \right)^{-1} \qwbF_{l\backslash k}^{(t)}$. Then we get
\begin{align}
    \qx_{l \rightarrow k}^{(t)}
    &\approx \left( \qwbE_{l}^{(t)} + \qI_{N_l} \right)^{-1} \qwbF_{l}^{(t)}
    - \left( \qwbE_{l}^{(t)} + \qI_{N_l} \right)^{-1} \qDelta \qwbF_{l \rightarrow k}^{(t)}
    \nonumber \\
    &= \qx_{l}^{(t)} - \left( \qwbE_{l}^{(t)} + \qI_{N_l} \right)^{-1} \qDelta \qwbF_{l \rightarrow k}^{(t)}. \label{eq:xx1}
\end{align}
We will approximate the above two terms by dropping their negligible components. Before proceeding, we deal with the approximation of $\qwbE_{l}^{(t)}$. Let us define $\qOmega_k^{(t)} \triangleq
\sum_{j} \qH_{k,j} \qV_{q_{j \rightarrow k}}^{(t-1)} \qH_{k,j}^H$. Then we have
\begin{align}
     \qwbE_{l}^{(t)}
     &= \sum_{k} \qH_{k,l}^H\left( \qOmega_k^{(t)} - \qH_{k,l} \qV_{l \rightarrow k}^{(t-1)} \qH_{k,l}^H + \beta \qI_{M_k} \right)^{-1}\qH_{k,l} \nonumber \\
    &\approx \sum_{k} \qH_{k,l}^H\left( \qOmega_k^{(t)} + \beta \qI_{M_k} \right)^{-1}\qH_{k,l}
    \triangleq \qSigma_{l}^{(t)}, \label{eq:sigma_final_ap}
\end{align}
where the approximation follows from the fact that $\qH_{k,l} \qV_{l \rightarrow k}^{(t-1)} \qH_{k,l}^H$ is of order $1/N_l$ and can be safely neglected. Similarly, we note that $\qV_{l
\rightarrow k}^{(t-1)}$ is nearly independent of $k$. This leads to
\begin{equation}
    \qV_{l \rightarrow k}^{(t-1)}
    = \left(\qwbE_{l\backslash k}^{(t-1)} + \qI_{N_l} \right)^{-1}
    \approx \left(\qSigma_{l}^{(t-1)} + \qI_{N_l} \right)^{-1} \triangleq \qV_{l}^{(t-1)}.
\end{equation}
Then we get
\begin{equation} \label{eq:Omega_fin}
    \qOmega_k^{(t)}= \sum_{j} \qH_{k,j} \qV_{q_{j \rightarrow k}}^{(t-1)} \qH_{k,j}^H
    \approx \sum_{j} \qH_{k,j} \qV_{j}^{(t-1)}  \qH_{k,j}^H.
\end{equation}

Now, we return to the approximation of $\qx_{l \rightarrow k}^{(t)}$. First, we deal with the second terms of (\ref{eq:xx1}) and get
\begin{align*}
 \qx_{l \rightarrow k}^{(t)}
 &\approx \qx_l^{(t)} - \left( \qwbE_{l}^{(t)} + \qI_{N_l} \right)^{-1} \left[ \qH_{k,l}^H \left( \qOmega_k^{(t)} - \qH_{k,l} \qV_{l \rightarrow k}^{(t-1)} \qH_{k,l}^H + \beta \qI_{M_k} \right)^{-1}
    \left( \qnu_k^{(t)} + \qH_{k,l} \qx_{l \rightarrow k}^{(t-1)} \right) \right] \\
 &\approx \qx_l^{(t)} -
 \left(\qSigma_{l}^{(t)} + \qI_{N_l} \right)^{-1} \qH_{k,l}^H \left( \qOmega_k^{(t)} + \beta \qI_{M_k} \right)^{-1} \qnu_k^{(t)},
\end{align*}
where the first approximation is directly from (\ref{eq:xx1}) by substituting the definition of $\qDelta \qwbF_{l \rightarrow k}^{(t)}$ and we have defined $\qnu_k^{(t)} \triangleq \qs_k^{(t-1)}
- \sum_{j} \qH_{k,j} \qx_{q_{j \rightarrow k}}^{(t-1)}$. Substituting the above approximation of $\qx_{l \rightarrow k}^{(t)}$ in $\qnu_k^{(t)}$, we get
\begin{align}
    \qnu_k^{(t)}
    &\approx \qs_k^{(t-1)} - \sum_{j} \qH_{k,j} \qx_j^{(t-1)} - \left( \sum_{j} \qH_{k,j} (\qSigma_{j}^{(t-1)} + \qI_{N_j} )^{-1}
    \qH_{k,j}^H \right) \left( \qOmega_k^{(t-1)} + \beta \qI_{M_k} \right)^{-1} \qnu_k^{(t-1)} \nonumber \\
    &= \qs_k^{(t-1)} - \sum_{j} \qH_{k,j} \qx_j^{(t-1)} - \qOmega_k^{(t)} \left( \qOmega_k^{(t-1)} + \beta \qI_{M_k} \right)^{-1} \qnu_k^{(t-1)}, \label{eq:nu_final_ap}
\end{align}
where the second equality follows from (\ref{eq:Omega_fin}).

Now, it remains to complete the calculation of $\qx_{l}^{(t)}$. We start from the definition
\begin{equation}
    \qx_{l}^{(t)} = \left[ \left(\qwbE_{l}^{(t)} \right)^{-1} + \qI_{N_l} \right]^{-1} \left(\qwbE_{l}^{(t)} \right)^{-1} \qwbF_{l}^{(t)}
    = \left[ \left(\qwbE_{l}^{(t)} \right)^{-1} + \qI_{N_l} \right]^{-1} \qmu_l^{(t)},
\end{equation}
where we have defined
\begin{equation}
    \qmu_l^{(t)} \triangleq \left( \qwbE_{l}^{(t)} \right)^{-1} \qwbF_{l}^{(t)}.
\end{equation}
Following the similar approximations as above, we get
\begin{align}
    \qmu_l^{(t)}
  = &
   ~\left[
    \sum_{k} \qH_{k,l}^H\left( \qOmega_k^{(t)} - \qH_{k,l} \qV_{l \rightarrow k}^{(t-1)} \qH_{k,l}^H + \beta \qI_{M_k} \right)^{-1}\qH_{k,l}
   \right]^{-1} \nonumber \\
   & \times
    \left[ \sum_{k} \qH_{k,l}^H \left( \qOmega_k^{(t)} - \qH_{k,l} \qV_{l \rightarrow k}^{(t-1)} \qH_{k,l}^H + \beta \qI_{M_k} \right)^{-1}
    \left( \qnu_k^{(t)} + \qH_{k,l} \qx_{l \rightarrow k}^{(t-1)} \right) \right] \nonumber \\
   \approx &  ~\qx_{l}^{(t-1)}
   + \left(\qSigma_{l}^{(t)}\right)^{-1}
   \left[ \sum_{k} \qH_{k,l}^H \left( \qOmega_k^{(t)} + \beta \qI_{M_k} \right)^{-1} \qnu_k^{(t)}  \right] \label{eq:mu_final_app}
\end{align}
and then
\begin{align} \label{eq:x_final_app}
    \qx_{l}^{(t)}
    \approx
    \left( \qSigma_{l}^{(t)} + \qI_{N_l} \right)^{-1} \qSigma_{l}^{(t)}\qmu_l^{(t)}.
\end{align}

Putting the above relations (\ref{eq:sigma_final_ap}), (\ref{eq:Omega_fin}), (\ref{eq:nu_final_ap}), (\ref{eq:mu_final_app}) and (\ref{eq:x_final_app}) together, we get AMP-RZFBF.

\section*{Appendix B: Lemmas}
For convenience, we provide some mathematical tools needed in this paper.

\begin{Lemma}\label{Lemma 1}
Given a positive definite matrix $\qA$, we have
\begin{equation}
    \frac{1}{\sfZ}\int \qx e^{-\qx^H\qA\qx + \qb^H \qx + \qx^H \qb } d\qx    = \qA^{-1} \qb,
\end{equation}
where $\sfZ$ is a normalization factor such that $1/\sfZ \int e^{-\qx^H\qA\qx + \qb^H \qx + \qx^H \qb } d\qx = 1$.
\end{Lemma}

\begin{Lemma}\label{Lemma 2}
A random matrix $\qX \in \bbC^{M \times N}$ is said to have a matrix variate complex Gaussian distribution with mean $\bar{\qX}$ and covariance matrix $\qB \otimes\qA$, if it can be written by
$\bar{\qX} + \qA^{\frac{1}{2}} \qW \qB^{\frac{1}{2}}$, where $\qA \in \bbC^{N \times N}$ and $\qB \in \bbC^{M \times M}$ are both positive definite and the elements of $\qW$ are i.i.d.~complex
Gaussian random variables with zero mean and unit variance. Then we have
\begin{align}
 \Ex\{ \qX\qC\qX^H \} &= \bar{\qX}\qC\bar{\qX}^H + \tr(\qB\qC) \qA, \\
 \Ex\{ \qX^H\qD\qX \} &= \bar{\qX}^H\qD\bar{\qX} + \tr(\qA\qD) \qB.
\end{align}
\end{Lemma}

{\renewcommand{\baselinestretch}{1.1}
\begin{footnotesize}
\bibliographystyle{IEEEtran}

\begin{thebibliography}{10}
\providecommand{\url}[1]{#1} \csname url@samestyle\endcsname \providecommand{\newblock}{\relax} \providecommand{\bibinfo}[2]{#2}
\providecommand{\BIBentrySTDinterwordspacing}{\spaceskip=0pt\relax} \providecommand{\BIBentryALTinterwordstretchfactor}{4}
\providecommand{\BIBentryALTinterwordspacing}{\spaceskip=\fontdimen2\font plus \BIBentryALTinterwordstretchfactor\fontdimen3\font minus
  \fontdimen4\font\relax}
\providecommand{\BIBforeignlanguage}[2]{{%
\expandafter\ifx\csname l@#1\endcsname\relax
\typeout{** WARNING: IEEEtran.bst: No hyphenation pattern has been}%
\typeout{** loaded for the language `#1'. Using the pattern for}%
\typeout{** the default language instead.}%
\else \language=\csname l@#1\endcsname \fi #2}} \providecommand{\BIBdecl}{\relax} \BIBdecl

\bibitem{Spencer-04COMMag}
{Q. H. Spencer, C. B. Peel, A. L. Swindlehurst, and M. Haardt}, ``{An
  introduction to the multiuser MIMO downlink},'' \emph{IEEE Commun. Mag.},
  vol.~42, no.~10, pp. 60--67, Oct. 2004.

\bibitem{ZGPan-04}
{G. Z. Pan, K. K. Wong, and T. S. Ng}, ``{Generalized multiuser orthogonal
  space division multiplexing},'' \emph{IEEE Trans. Wireless Commun.}, vol.~3,
  no.~6, pp. 1--5, Nov. 2004.

\bibitem{Gesbert-07SigMag}
{D. Gesbert, M. Kountouris, R. W. Heath Jr., C.-B. Chae, and T. Salzer},
  ``{Shifting the MIMO paradigm},'' \emph{IEEE Sig. Proc. Mag.}, vol.~24,
  no.~5, pp. 36--46, Sep. 2007.

\bibitem{Shamai-97IT}
{S. Shamai (Shitz) and A. D. Wyner}, ``{Information-theoretic considerations
  for symmetric, cellular, multiple-access fading channels -- Part I \& II},''
  \emph{IEEE Trans. Inf. Theory}, vol.~43, no.~6, pp. 1877--1911, Nov. 1997.

\bibitem{Caire-03IT}
{G. Caire and S. Shamai (Shitz)}, ``{On the achievable throughput of a
  multiantenna Gaussian broadcast channel},'' \emph{IEEE Trans. Inf. Theory},
  vol.~49, no.~7, pp. 1691--1706, July 2003.

\bibitem{Somekh07IT}
{O.~Somekh, B.~M.~Zaidel, and S.~Shamai (Shitz)}, ``Sum rate characterization
  of joint multiple cell-site processing,'' \emph{{IEEE} Trans. Info. Theory},
  vol.~53, no.~12, pp. 4473--4497, Dec. 2007.

\bibitem{Somekh09IT}
{O.~Somekh, O.~Simeone, Y.~Bar-Ness, A.~M.~Haimovich, and S.~Shamai (Shitz)},
  ``{Cooperative multicell zero-forcing beamforming in cellular downlink
  channels},'' \emph{{IEEE} Trans. Info. Theory}, vol. 2009, no.~7, pp.
  3206--3219, Jul. 2009.

\bibitem{Gesbert10JSAC}
{D. Gesbert, S. Hanly, H. Huang, S. Shamai, O. Simeone, and W. Yu},
  ``{Multi-cell MIMO cooperative networks: A new look at interference},''
  \emph{IEEE J. Sel. Areas Commun. Special Issue Cooperative Commun. MIMO
  Cellular Net.}, vol.~28, no.~9, pp. 1--29, Dec. 2010.

\bibitem{Akai-Kit-10}
{C. K. Wen and K. K. Wong}, ``{On the sum-rate of uplink MIMO cellular systems
  with amplify-and-forward relaying and collaborative base stations},''
  \emph{IEEE J. Sel. Areas Commun. Special Issue Cooperative Commun. MIMO
  Cellular Net.}, vol.~28, no.~9, pp. 1409--1424, Dec. 2010.

\bibitem{Costa-83IT}
{M. H. M. Costa}, ``{Writing on dirty paper},'' \emph{IEEE Trans. Inf. Theory},
  vol.~29, no.~3, pp. 439--441, May 1983.

\bibitem{Yoo-06JSAC}
{T. Yoo and A. Goldsmith}, ``{On the optimality of multiantenna broadcast
  scheduling using zero-forcing beamforming},'' \emph{IEEE J. Sel. Areas
  Commun.}, vol.~24, no.~3, pp. 528--541, Mar. 2006.

\bibitem{Sam-07}
{D. Samardzija, H. Huang, T. Sizer, and R. Valenzuela}, ``{Experimental
  downlink multiuser MIMO system with distributed and coherentlycoordinated
  transmit antennas},'' in \emph{Proc. Int. Conference on Communications
  (ICC'09)}, Jun. 2007.

\bibitem{Irm-09}
{R. Irmer, H.-P. Mayer, A. Weber, V. Braun, M. Schmidt, M. Ohm, N. Ahr, A.
  Zoch, C. Jandura, P. Marsch, and G. Fettweis}, ``{Multisite field trial for
  LTE and advanced concepts},'' \emph{IEEE Commun. Mag.}, vol.~47, no.~2, pp.
  92--98, Feb. 2009.

\bibitem{Jun-10}
{V. Jungnickel, A. Forck, S. Jaeckel, F. Bauermeister, S. Schiffermueller, S.
  Schubert, S. Wahls, L. Thiele, T. Haustein, W. Kreher, J. Mueller, H. Droste,
  and G. Kadel}, ``{Field trials using coordinated multi-point transmission in
  the downlink},'' in \emph{Proc. 3rd Int. Workshop on Wireless Distributed
  Networks (WDN)}, 2010.

\bibitem{Irm-11}
{R. Irmer, H. Droste, P. Marsch, M. Grieger, G. Fettweis, S. Brueck, H.-P.
  Mayer, L. Thiele, and V. Jungnickel}, ``{Coordinated multipoint: Concepts,
  performance, and field trial results},'' \emph{IEEE Commun. Mag.}, vol.~49,
  no.~2, pp. 102--111, Feb. 2011.

\bibitem{Joham-02ISSSTA}
{M. Joham, K. Kusume, M. H. Gzara, W. Utschick, and J. A. Nossek}, ``{Transmit
  Wiener filter for the downlink of TDDDS-CDMA systems},'' in \emph{Proc. IEEE
  7th Int. Symp. Spread-Spectrum Tech. Appl. (ISSSTA)}, vol.~1, 2002, pp.
  9--13.

\bibitem{Peel-05Tcom}
{C. B. Peel, B. M. Hochwald, and A. L. Swindlehurst}, ``{A vector-perturbation
  technique for near-capacity multiantenna multiuser communication--Part I:
  channel inversion and regularization},'' \emph{IEEE Trans. Commun.}, vol.~53,
  no.~1, pp. 195--202, Jan. 2005.

\bibitem{Zak-12IT}
{R. Zakhour and S. Hanly}, ``{Base station cooperation on the downlink: Large
  system analysis},'' \emph{IEEE Trans. Inf. Theory}, vol.~58, no.~4, pp.
  2079--2106, Apr. 2012.

\bibitem{Simeone-11BOOK}
{O. Simeone, N. Levy, A. Sanderovich, O. Somekh, B. M. Zaidel, H. V. Poor, and
  S. Shamai (Shitz)}, \emph{{Cooperative wireless cellular systems: An
  information-theoretic view}}.\hskip 1em plus 0.5em minus 0.4em\relax
  Foundations and Trends in Commun. Inf. Theory, 2011.

\bibitem{Hoydis-12JSAC}
{J. Hoydis, S. ten Brink, and M. Debbah}, ``{Massive MIMO in the UL/DL of
  cellular networks: How many antennas do we need?}'' \emph{IEEE J. Sel. Areas
  Commun.}, vol.~31, no.~2, pp. 160--171, Feb. 2013.

\bibitem{Ng-08IT}
{B. L. Ng, J. S. Evans, S. V. Hanly, and D. Aktas}, ``{Distributed downlink
  beamforming with cooperative base stations},'' \emph{IEEE Trans. Inf.
  Theory}, vol.~54, no.~12, pp. 5491--5499, Dec. 2008.

\bibitem{Aktas-11BOOK}
{E. Aktas, D. Aktas, S. V. Hanly and J. S. Evans}, \emph{{Turbo Base
  Stations}}.\hskip 1em plus 0.5em minus 0.4em\relax in Cooperative Cellular
  Wireless Networks (E. Hossain, D. In Kim and V. K. Bhargava, eds.), Cambridge
  University Press, 2011.

\bibitem{Bjornson-10SP}
{E. Bj\"{o}rnson, R. Zakhour, D. Gesbert, and B. Ottersten}, ``{Cooperative
  multicell precoding: Rate region characterization and distributed strategies
  with instantaneous and statistical CSI},'' \emph{IEEE Trans. Sig. Proc.},
  vol.~58, no.~8, pp. 4298--4310, Aug. 2010.

\bibitem{Sohn-11TWC}
{I. Sohn, S. H. Lee, and J. G. Andrews}, ``{Belief propagation for distributed
  downlink beamforming in cooperative MIMO cellular networks},'' \emph{IEEE
  Trans. Wireless Commun.}, vol.~12, no.~10, pp. 4140--4149, Dec. 2011.

\bibitem{Huang-11SP}
{Y. Huang, G. Zheng, M. Bengtsson, K.-K. Wong, L. Yang, and B. Ottersten},
  ``{Distributed multicell beamforming with limited intercell coordination},''
  \emph{IEEE Trans. Sig. Proc.}, vol.~59, no.~2, pp. 728--738, Feb. 2011.

\bibitem{Rangan-12JSAC}
{S. Rangan and R. Madan}, ``{Belief propagation methods for intercell
  interference coordination in femtocell networks},'' \emph{IEEE J. Sel. Areas
  Commun.}, vol.~30, no.~3, pp. 631--640, Apr. 2012.

\bibitem{Donoho-09PNAS}
{D. L. Donoho, A. Maleki, and A. Montanari}, ``{Message passing algorithms for
  compressed sensing},'' \emph{Proceedings of the National Academy of
  Sciences}, 2009.

\bibitem{Bayati-11IT}
{M. Bayati and A. Montanari}, ``{The dynamics of message passing on dense
  graphs, with applications to compressed sensing,},'' \emph{IEEE Trans. Inf.
  Theory}, vol.~57, no.~2, pp. 764--785, Feb. 2011.

\bibitem{Krzakala-12JSM}
{F. Krzakala, M. M\'{e}zard, F. Sausset, Y. Sun, and L. Zdeborov\'{a}},
  ``{Probabilistic reconstruction in compressed sensing: algorithms, phase
  diagrams, and threshold achieving matrices},'' \emph{J. Stat. Mech.}, vol.
  P08009, 2012.

\bibitem{Rangan-10ArXiv}
{S. Rangan}, ``{Generalized approximate message passing for estimation with
  random linear mixing},'' \emph{ArXiv e-prints:1010.5141v2 [cs.IT]}, 2010.

\bibitem{Hoydis-11ISIT}
{J. Hoydis, M. Debbah, and M. Kobayashi}, ``{Asymptotic moments for
  interference mitigation in correlated fading channels},'' in \emph{IEEE
  Internatinal Sym. Inf. Theory (ISIT'11)}, Saint-Petersburg, Russia, 2011.

\bibitem{Lakshminarayana-12SP}
{S. Lakshminarayana, M. Debbah, and M. Assaad}, ``{Asymptotic analysis of
  distributed multi-cell beamforming},'' \emph{IEEE Trans. Sig. Proc.}, 2012,
  submitted.

\bibitem{Boyd-11BOOK}
S.~Boyd, N.~Parikh, E.~Chu, B.~Peleato, and J.~Eckstein, \emph{Distributed
  Optimization and Statistical Learning via the Alternating Direction Method of
  Multipliers}.\hskip 1em plus 0.5em minus 0.4em\relax Foundations and Trends
  in Machine Learning, 2011.

\bibitem{Zhu-10TWC}
{H. Zhu, A. Cano, and G. Giannakis}, ``{Distributed consensus-based
  demodulation: algorithms and error analysis},'' \emph{IEEE Trans. Wireless
  Commun.}, vol.~9, no.~6, pp. 2044--2054, June 2010.

\bibitem{Shiu-00TCOM}
{D. Shiu, G. J. Foschini, M. J. Gans, and J. M. Kahn}, ``{Fading correlation
  and its effect on the capacity of multi-element antenna systems},''
  \emph{IEEE Trans. Commun.}, vol.~48, no.~3, pp. 502--513, Mar. 2000.

\bibitem{Poor-94BOOK}
H.~V. Poor, \emph{{An Introduction to Signal Detection and Estimation}}.\hskip
  1em plus 0.5em minus 0.4em\relax New York: Springer-Verlag, 1994.

\bibitem{Kschischang-01IT}
{F. R. Kschischang, B. J. Frey, and H.-A. Loeliger}, ``{Factor graphs and the
  sum-product algorithm},'' \emph{IEEE Trans. Inf. Theory}, vol.~47, no.~2, pp.
  498--519, Feb. 2001.

\bibitem{Guo-06ITW}
{D. Guo and C.-C. Wang}, ``{Asymptotic mean-square optimality of belief
  propagation for sparse linear systems},'' in \emph{in Proc. IEEE Inform.
  Theory Workshop}, Chengdu, China, Oct. 2006, pp. 194--198.

\bibitem{Bai-10}
Z.~Bai and J.~W. Silverstein, \emph{Spectral Analysis of Large Dimensional
  Random Matrices}.\hskip 1em plus 0.5em minus 0.4em\relax Springer Series in
  Statistics, 2010.

\end{thebibliography}

\end{footnotesize}}
\end{document}